\documentclass[preprint,showpacs,preprintnumbers,amsmath,amssymb]{revtex4}

\usepackage{graphicx}
\usepackage{dcolumn}
\usepackage{bm}

\begin{document}

\title{Fourfold oscillations and anomalous magnetic irreversibility of magnetoresistance 
in the non-metallic regime of Pr$_{1.85}$Ce$_{0.15}$CuO$_4$}
%
\author{P. Fournier, M.-E. Gosselin, S. Savard, J. Renaud, I. Hetel, P. Richard and G. Riou}
\affiliation{D\'epartement de Physique, Universit\'e de Sherbrooke, 
Sherbrooke, Qu\'ebec, CANADA, J1K 2R1}
\date{\today}
%
\begin{abstract}
Using magnetoresistance measurements as a function of applied 
magnetic field and its direction of application, 
we present sharp angular-dependent magnetoresistance oscillations for the electron-doped cuprates in their 
low-temperature non-metallic regime. The presence of irreversibility in the magnetoresistance 
measurements and the related strong anisotropy of the field dependence for different in-plane magnetic 
field orientations indicate that magnetic domains play an important role for the determination 
of electronic properties. These domains are likely related to the stripe phase reported previously in 
hole-doped cuprates.
\end{abstract}

\maketitle

%
\eject
%
%
\par
In the low doping regime of high temperature superconductors (HTSC), 
resistivity crosses over most often from a high temperature metallic-like state to a low temperature 
non-metallic behavior\cite{crossover}.  Interestingly, the crossover temperature observed by in-plane 
resistivity {\it does not} correspond to the N\'eel temperature ($T_N$), the onset of long range 
antiferromagnetic order. In fact, there is no hint of this transition 
in $\rho_{xx}(T)$ as metallic-like behavior persists sometimes {\it below} $T_N$ \cite{rhoxx-Tn1,rhoxx-Tn2}.  
Ando and coworkers argued that this independence of $\rho_{xx}(T)$ from the underlying 
magnetic order is a sign of phase separation, probably with conducting stripes 
separated by antiferromagnetic domains. Such self-organized structures of 
the charge carriers was shown to lead to intriguing behaviors for several physical properties\cite{rhoxx-YBCO,rhoxx-LSCO,mag-aniso,kappa-aniso}. An anisotropic twofold response to 
magnetic field applied along the copper-oxygen (CuO$_2$) planes was reported 
for instance using in-plane magnetoresistance (MR) in the non-metallic regime 
of YBa$_2$Cu$_3$O$_{6.33}$ (YBCO)\cite{rhoxx-YBCO} and in lightly 
doped La$_{1.99}$Sr$_{0.01}$CuO$_4$ (LSCO)\cite{rhoxx-LSCO}.
\par
To avoid the possible contributions of orthorhombic distortions or phases (as in YBCO and LSCO), we have 
chosen the tetragonal electron-doped cuprates\cite{reviews} as good candidates to make a definitive test of 
anisotropic transport properties.  
We focus on the results obtained for magnetic fields always applied \textit{parallel} to 
the CuO$_2$ planes. In Figure 1, we present an example of raw 
in-plane and out-of-plane resistance data as a function of angle with respect to the in-plane 
crystal a-axis. These data were measured deep into the non-metallic regime of  
\textit{non-superconducting} Pr$_{1.85}$Ce$_{0.15}$CuO$_4$ (PCCO). We observe  
clear \textit{fourfold} and \textit{symmetrical} oscillations of the resistance with sharp maxima [minima] for $R_{xx}$ [$R_{c}$] for  
fields applied along the in-plane crystal axis (i.e. along the CuO bonds of the CuO$_2$ 
planes) and broad minima [maxima] for field applied along the diagonals. This effect can be 
emphasized using the polar plot of Fig. 1(c). 
\par
In this Letter, we present this magnetoresistance anisotropy, for 
current along the CuO$_2$ planes {\it and} along the c-axis (thickness) of 
non-superconducting electron-doped single crystals. We 
show for the first time that sharp fourfold oscillations persist over the whole non-metallic 
regime for both in-plane ($\rho_{xx}$) and out-of-plane ($\rho_c$) resistivities. 
More importantly, the field dependence of $\rho_c$ presents irreversibilities which cannot be 
explained by conventional band theory. We ascribe the new transport signatures to the presence of magnetic 
domains. Their origin is probably related to the presence of stripe domains (order) in the cuprates. 
%
%
%
\begin{figure}
	\begin{center}
\includegraphics[width=10cm, angle=-90]{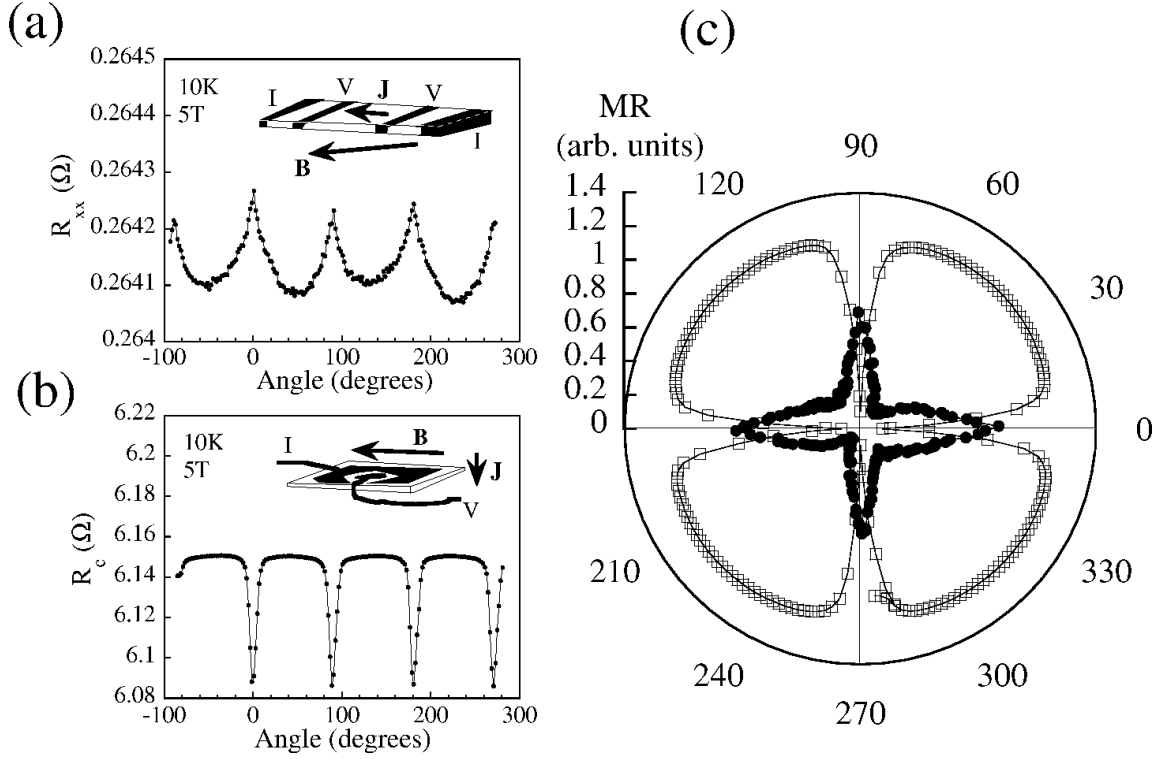}
	\end{center}
	\caption{Raw resistance data as a function of angle for non-superconducting Pr$_{1.85}$Ce$_{0.15}$CuO$_4$ : 
	(a) in-plane resistance $R_{xx}$ , (b) out-of-plane resistance $R_c$ and (c) Polar plot of the same 
	data. 0$^o$ corresponds to the field applied along the Cu-O bonds. The inserts in (a) and (b) show the 
	contact configurations used for the measurements.}
	\label{fig1}
\end{figure}
%
\par
%
\par
The single crystals of Re$_{2-x}$Ce$_x$CuO$_4$ (Re = Pr, Nd, Sm and Eu) used 
for this study were grown by the directional flux growth method in alumina and 
high purity magnesia crucibles\cite{xtal-ntype,MgO}. The as-grown crystals are known to be 
non-superconducting even for x = 0.15, and only a reduction process allows them 
to become superconducting with a maximum transition temperature between 18 and 25K 
depending on the rare-earth ion\cite{reviews}. In the present work, we study only as-grown 
{\it non-superconducting} crystals with x = 0.15 in order to extend as much 
as possible the temperature range for which a non-metallic regime is observed. 
\par
Silver epoxy contacts were applied directly onto the as-grown crystals (typical in-plane size : 2mm x 1 mm, 
and 30 $\mu m$ thickness along the c-axis) in two different configurations insuring uniform current 
density for the measurement of the in-plane ($\rho _{xx}$) and c-axis ($\rho _{c}$) 
resistivity as shown in the inserts of Figs. 1(a) and (b). The samples were 
then mounted on sapphire supports and attached onto a specially designed 
Physical Property Measurement System (PPMS) rotator chip, such that the applied 
magnetic field can be rotated over a full 360$^{\circ}$. 
The direction of the in-plane crystal axis (a-axis direction relative to the edge of the 
rectangular crystals) were determined using Laue x-ray
diffraction and by Raman scattering. Several 
tests were performed to rule out contact configuration, thermometry and misorientation problems. 
%
%
\begin{figure}
\includegraphics[width=10cm, angle=-90]{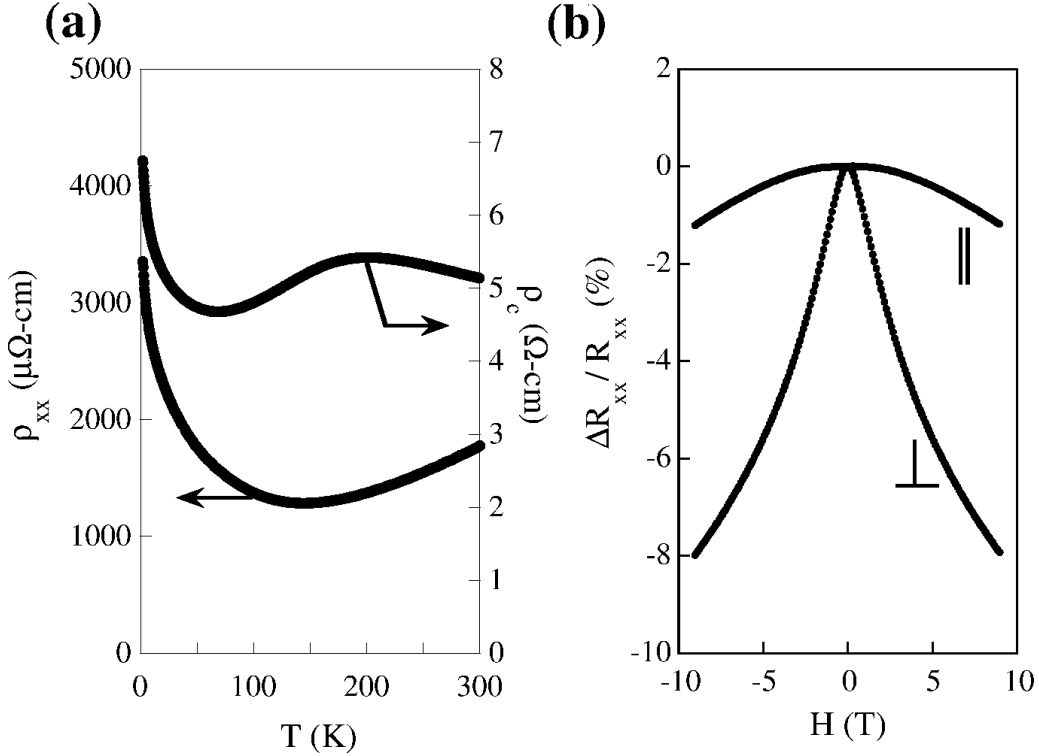}
\caption{(a) In-plane resistivity and c-axis resistivitity as a function of 
temperature for \textit{non-superconducting} Pr$_{1.85}$Ce$_{0.15}$CuO$_{4+\delta}$ crystals.
(b) in-plane magnetoresistance as a function of 
field for T = 5K and its anisotropy for field applied along the c-axis ($\perp$) 
and along the CuO$_2$ planes($\parallel$).}
\label{fig2}
\end{figure}
%
%
%
%
\par
In Figure 2(a), we present the typical temperature dependence of in-plane 
and c-axis resistivity components for \textit{non-superconducting} Pr$_{1.85}$Ce$_{0.15}$CuO$_{4}$ 
single crystals. Both components show a non-metallic regime at low 
temperatures where the MR oscillations are observed. The in-plane resistivity $\rho_{xx}$ is metallic-like 
at room temperature while it shows an upturn starting at $T_{min,xx} \approx 125K$   
which is very sensitive to the oxygen content. Below $T_{min,xx}$, 
$\rho_{xx}$ approaches a $\ln (T)$ behavior (not shown), similar to that reported 
for non-superconducting NCCO and PCCO\cite{saturation}. This temperature behavior and the strong anisotropy 
illustrated in Fig. 2(b) have been interpreted as a signature of two-dimensional weak localization (2DWL) 
by disorder\cite{2DWL}. 
For the c-axis resistivity $\rho_{c}$ , a non-metallic trend is observed at 
room temperature for several crystals, followed by a maximum (at 
$T \approx 200K$)\cite{comm2}, then by a metallic-like regime. It is finally followed by a non-metallic regime, 
also approaching $~\ln (T)$ , below $T_{min,c} \approx 70K$ .%
%
%
%
\begin{figure}
\includegraphics[width=10cm, angle=-90]{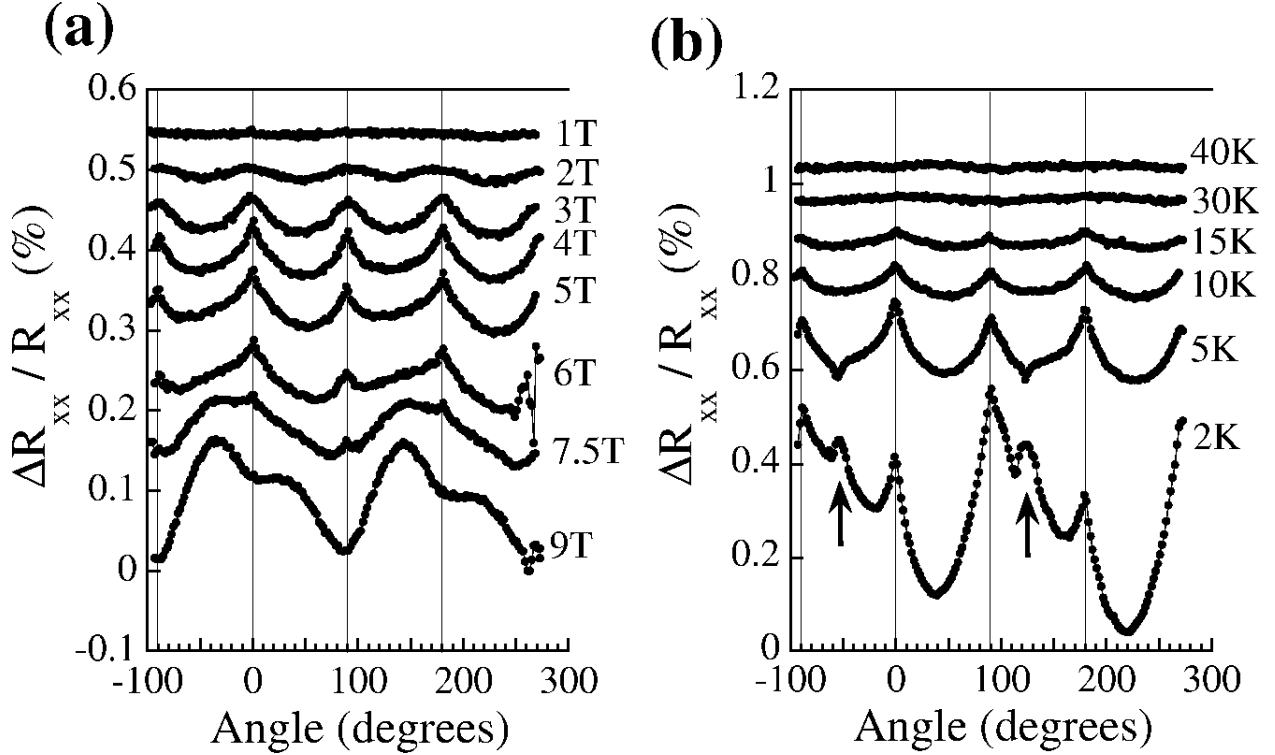}
\caption{In-plane magnetoresistance as a function of angle (a) for several fields at 10K; (b) for several temperatures at 5T. The data has been moved vertically for clarity.}
\label{fig3}
\end{figure}
%
\par
The field dependence of the in-plane 
magnetoresistance oscillations at 10K are presented in Fig. 3(a). The MR oscillations evolve steadily from broad 
features at 2T to sharp maxima for 4T whenever the field is applied along the Cu-O bonds.  
For fields approaching 9T, new maxima develop for fields applied approximately along 
the diagonal directions. The relative proportion of both oscillations 
varies also steadily with temperature as evidenced by the same 45$^{\circ}$ features barely appearing 
at 5T and 2K in Fig. 3(b) (see arrows). 
The amplitude of these oscillations in $\rho_{xx}$ represents only a small fraction 
($\Delta \rho_{osc} / \rho \sim 0.05 \% $) of the total negative MR [$\Delta \rho_{tot} / \rho \sim 1 \% $ : 
see Fig. 2(b)]. Thus, unlike YBCO and LSCO, 
$\rho_{xx}$ oscillations are very weak \textit{and} have almost perfect fourfold symmetry. 
\par
In Fig. 3(b), we present the evolution of these $\rho_{xx}$ oscillations (at 5T) with temperature. A similar 
effect is obtained if one decreases the temperature or increases the applied field. We observe that the 
weak oscillations disappear quickly with increasing temperature, vanishing close to $T_{min,xx}$. This seems 
to imply that this anisotropic behavior can only be observed in the non-metallic regime 
of these materials (within our sensitivity). 
%
%
%
%
\begin{figure}
\includegraphics[width=10cm, angle=-90]{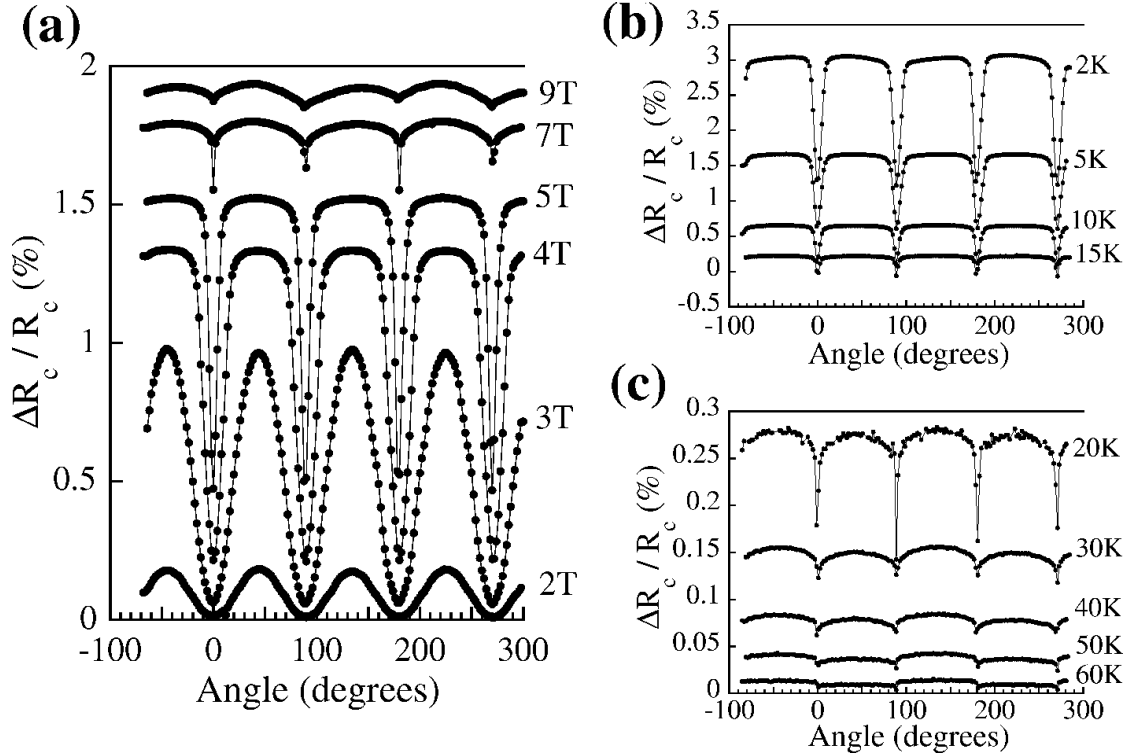}
\caption{Out-of-plane magnetoresistance as a function of angle: (a) for several fields at 5K; (b) and (c) for several temperatures at 5T. The data has been moved vertically for clarity.}
\label{fig4}
\end{figure}
%
%
\par
In Fig. 4(a), we present 
the angular dependence of $\rho_c$ in several magnetic fields at 5K. We observe very sharp minima in the c-axis resistivity, developing particularly for intermediate fields (3 to 6T). These sharp cusps appear for magnetic fields applied along the Cu-O bonds. As the applied magnetic field is further increased, these anomalies are gradually replaced by oscillations of smaller amplitude. In Figs. 4(b) and (c), we illustrate the strong temperature dependence of these features as they seem once again to vanish as the sample reaches temperatures close 
to the crossover to the metallic-like state. As will be shown below, the magnitude of the oscillations observed 
for $\rho_c$ are comparable to the total \textit{positive} MR, in sharp contrast with those observed with $\rho_{xx}$. Therefore, they are less sensitive to mis-orientation of the crystals and thus easily observed.

%
%
\begin{figure}
\includegraphics[width=10cm, angle=-90]{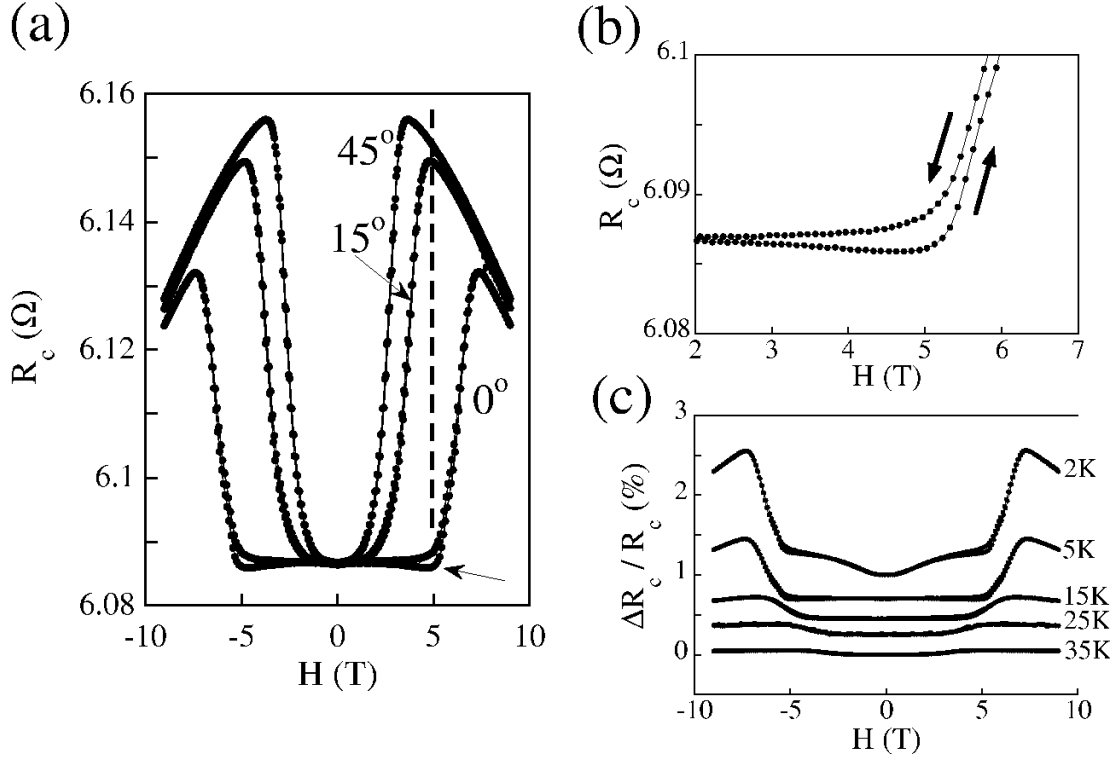}
\caption{Out-of-plane magnetoresistance as a function of field : (a) for 0, 15 and 45$^{\circ}$ at 5K; 
(b) same data magnified around 5T (arrows indicate field sweeping rate); and (c) for several temperatures 
for field along CuO bonds ($\theta = 0^{\circ}$). The data has been moved vertically for clarity.}
\label{fig5}
\end{figure}
%
%
\par
In Figure 5 (a), we present the c-axis magnetoresistance at T = 5K as a function of in-plane magnetic 
field applied along three different directions (0, 15 and 45$^{\circ}$). Contrary to the in-plane 
resistivity, the c-axis magnetoresistance is \textit{positive} and 
presents an unusual field dependence. For the field applied along 
the a-axis (at 0$^{\circ}$), the resistivity remains remarkably flat at low fields until a threshold field 
is reached. At this point, the resistance varies sharply, as if the system was crossing a transition. Beyond 
this threshold, the magnetoresistance resumes a high field $-B^2$ behavior. As soon as the field direction 
deviates from the a-axis [for 15 and 45$^\circ$ in Fig. 5(a)], the magnetoresistance presents a very sharp 
positive increase at low fields, quickly reaching the saturation $-B^2$ 
regime. The large magnitude of the oscillations in $\rho_c$ at about 4 - 5T in Fig. 4 (a) can easily be 
explained by the strong variations in resistance at 5T for different field orientations underlined by the 
dashed line in Fig. 5(a).  We should emphasize here that $\rho_{xx}$ displays its sharpest peaks in the same 
range of applied field.
\par
Interestingly, upon decreasing the applied magnetic field, the resistance for $\theta = 0^{\circ}$ 
shows a similar transition-like pattern, but the \textit{resistance presents a clear sign of irreversibility} 
observed in Fig. 5(b). The arising of magnetic hysteresis is a clear demonstration of the 
influence of magnetic history, i.e. the presence of magnetic domains. These domains (and their  
domain walls) are strongly affected by the direction and strength of the 
in-plane rotating field, and in turn 
they affect the conductivity\cite{comm4}. We can interpret the data in Fig. 5(a) as evidence 
of easy magnetization axis along the diagonal directions, while the Cu-O bond directions (a and "b" axis) 
correspond to hard axis. Since the c-axis resistivity is a measure of interplane tunneling  
between CuO$_2$ planes\cite{interplane}, the alignment of magnetic 
domains at high fields is \textit{detrimental} to interplane tunneling causing an increase of 
c-axis resistivity. This effect is present up to temperatures approaching $T_{min,c}$ [see Fig. 5(c)]. 
Inversely, in-plane resistivity $\rho_{xx}$ presents peaks for fields  
applied along the hard axis and flat minima for field applied along the diagonal easy axis. 
\par
The behavior of the magnetoresistance anisotropy for PCCO is significantly  
different from YBCO\cite{rhoxx-YBCO} and LSCO\cite{rhoxx-LSCO}. 
For YBCO\cite{rhoxx-YBCO}, there is a clear sign change when magnetic field is applied 
parallel and transverse to the applied in-plane current. In LSCO, no sign change, but the presence of four 
asymmetrical lobes was attributed to twinning. In our case, we never observe a sign change in the MR with angle (it always remains negative) and the oscillations are not sensitive to the direction of the 
applied current as they remain about the same height for the field parallel [0$^\circ$] and 
transverse [90$^\circ$] to the current, except for the additional features observed at high magnetic fields in $\rho_{xx}$. The magnitude of the in-plane MR oscillations is much smaller for PCCO  
than for YBCO and LSCO. In both cases however, the oscillations disappear close the non-metal to metal 
crossover. We suspect that orthorhombic distortions could be a key player in promoting the differences in the 
magnitude and the anisotropy of these MR oscillations in YBCO, LSCO and PCCO.
\par
MR oscillations 
of c-axis resistivity in cuprates were first reported in strongly overdoped Tl-based cuprates\cite{Hussey-MR} 
with $T_c \approx 25K$. In this case, the fourfold oscillations are large (0.33$\%$ of the total resistance at 
~40K and 10T) and correspond to a doping region of the phase diagram where both $\rho_{xx}$ and $\rho_{c}$ remain metallic-like over the whole temperature range. Because the mean-free path (MFP) is fairly large in this 
strongly overdoped regime, Dragulescu et al. argued that the oscillations in Tl-based cuprates could be due 
to angular-dependent magnetoresistance oscillations\cite{Dragulescu}. In our case, the very 
small MFP for this doping\cite{PF-Nernst,Gollnik} precludes such interpretation. 
\par
Recent experiments indicate that the 
carriers injected into the cuprates through chemical substitution have a strong tendency to distribute  
non-uniformly in the copper-oxygen planes, segregating into phase-separated 
regions, clusters and stripes\cite{Carlson}. Several experimental observations, including the MR oscillations 
obtained with YBCO\cite{rhoxx-YBCO} and LSCO\cite{rhoxx-YBCO}, fit into this possible 
scenario. However, only a recent report by Sun \textit{et al.} indicates the possible presence of 
stripes in the electron-doped cuprates\cite{Ando-PLCCO}. Assuming  
their existence in the electron-doped 
cuprates\cite{comm6}, we should expect the c-axis resistivity to decrease whenever stripes in adjacent CuO$_2$ planes are directly on top of each other and aligned in the same direction : in this particular case, 
electrons can tunnel more easily between planes, thus decreasing resistivity. However, our data show that 
$\rho_c$ is higher when the field is high (for well aligned domains). We suggest here that a low 
density of stripes in adjacent planes makes it more difficult to have stripes on top of each other 
whenever domains are well aligned (high fields), in particular if domains on adjacent planes 
are not or weakly correlated (as in LSCO\cite{Matsuda}). At low fields, random orientation of 
stripe domains could lead to a better overlap from weakly correlated adjacent planes, and thus to a better conductivity. For in-plane MR [see Fig. 1(a)], the application of in-plane magnetic field could  
promote a partial displacement of domain walls (the stripes)\cite{fl-stripes}, enough to change 
the resistivity and improve the channeling of electrons along longer conducting "rivers of charges". 
Because the in-plane MR oscillations are so small, the changes in domain wall configurations are probably scarse.  
Our data would suggest that such modifications of the wall configuration are hard to develop only 
along the hard axis (the Cu-O bonds), leading to maxima in $\rho_{xx}$.
\par
We should mention that this simple scenario ignores completely the possibility of interaction 
with the underlying rare earth magnetic 
order (different in PrCeCuO, NdCeCuO, SmCeCuO)\cite{Hundley}. Our preliminary data on 
NdCeCuO and SmCeCuO showed no significant difference for the field and angular dependence of $\rho_{ab}$ and 
$\rho_c$ , thus ruling out a direct implication of rare earth magnetism. Moreover, it remains unclear 
how an electronic system could present at the same time signatures consistent with 2DWL by disorder and one 
dimensional features like stripes, unless the spin stripes induce only a partial charge density wave in 
the CuO$_2$ planes. This aspect will need further exploration, possibly through the doping dependence of 
the observed oscillations.  
%
\par
In summary, we presented anomalous magnetoresistance oscillations for non-superconducting electron-doped 
cuprates in their non-metallic regime for magnetic field applied along the copper-oxygen planes.  We 
showed that sharp fourfold oscillations persist over the whole non-metallic 
regime for both in-plane ($\rho_{xx}$) and out-of-plane ($\rho_c$) resistivities. 
The field dependence of $\rho_c$ presents irreversibilities which can be 
explained by the presence of magnetic domains. Their origin is probably related to the 
presence of stripe domains in the cuprates. 
%
\par
We thank A.-M. Tremblay, C. Bourbonnais, K. LeHur and M. Li for several discussions and S. Pelletier for 
technical assistance. We acknowledge the support of CIAR, CFI, NSERC and the Fondation of 
the Universit\'e de Sherbrooke.  
%
\bibliography{Fournier}

\begin{thebibliography}{25}
\bibitem[{com({\natexlab{b}})}]{comm4}
\bibinfo{note}{In all the measurements done on in-plane magnetoresistance, we
  never observed the irreversibilities, which could be due to the small
  magnitude of the MR for $\rho_{xx}$ and our limited sensitivity ($\sim
  1/20000$).}
\end{thebibliography}
\end{document}